\documentclass[aps,prb,twocolumn,groupedaddres]{revtex4}
\usepackage[dvips]{graphics}
\usepackage{bm}
\begin{document} 
\title{Formalism for obtaining nuclear momentum distributions by the Deep 
  Inelastic Neutron Scattering technique}
\author{J.J. Blostein}
\affiliation{Consejo Nacional de
  Investigaciones Cient{\'\i}ficas y T{\'e}cnicas, Centro At{\'o}mico Bariloche
  and Instituto Balseiro, Comisi{\'o}n Nacional de E\-ner\-g{\'\i}\-a At{\'o}mica,
  Universidad Nacional de Cuyo,(8400) Bariloche, Argentina}          
\author{J.  Dawidowski}     
\affiliation{Consejo Nacional de
  Investigaciones Cient{\'\i}ficas y T{\'e}cnicas, Centro At{\'o}mico Bariloche
  and Instituto Balseiro, Comisi{\'o}n Nacional de E\-ner\-g{\'\i}\-a At{\'o}mica,
  Universidad Nacional de Cuyo,(8400) Bariloche, Argentina}          
\email{javier@cab.cnea.gov.ar}
\author{J.R. Granada}     
\affiliation{Consejo Nacional de
  Investigaciones Cient{\'\i}ficas y T{\'e}cnicas, Centro At{\'o}mico Bariloche
  and Instituto Balseiro, Comisi{\'o}n Nacional de E\-ner\-g{\'\i}\-a At{\'o}mica,
  Universidad Nacional de Cuyo,(8400) Bariloche, Argentina}          
\date{\today}

\begin{abstract}
  We present a new formalism to obtain momentum distributions in
  condensed matter from Neutron Compton Profiles measured by the Deep
  Inelastic Neutron Scattering technique. The formalism describes
  exactly the Neutron Compton Profiles as an integral in the momentum
  variable $y$. As a result we obtain a Volterra equation of the first
  kind that relates the experimentally measured magnitude with the
  momentum distributions of the nuclei in the sample. The integration
  kernel is related with the incident neutron spectrum, the total
  cross section of the filter analyzer and the detectors efficiency
  function.  A comparison of the present formalism with the
  customarily employed approximation based on a convolution of the
  momentum distribution with a resolution function is presented. We
  describe the inaccuracies that the use of this approximation
  produces, and propose a new data treatment procedure based on the
  present formalism.

\end{abstract}
\maketitle                                                                      

\section{Introduction} 

During the last decades there has been a growing interest in the
experimental determination of momentum distributions of different
particles in a wide variety of systems, being of principal importance
the experiments based on deep-inelastic scattering techniques. Thus,
the Compton scattering of x rays, gamma rays, and electrons has been
for a long time used to obtain the electron momentum distribution in
condensed matter, the momentum distributions of nucleons in nuclei have
been determined by inelastic scattering of high-energy protons and
electrons, and the deep inelastic scattering of electrons, muons and
neutrinos was essential for establishing the quark model of the
nucleons \cite{Sears84, Breidenbach}.

The experimental determination of atomic momentum distribution in
condensed matter making use of different neutron techniques has
experienced a considerable evolution up to the present times. An
historical review of these neutron experimental techniques has been
presented in Ref. \onlinecite{Holt}. Notably, the Deep Inelastic
Neutron Scattering (DINS) technique, proposed by Hohenberg and
Platzmann \cite{Hohenberg} nearly 40 years ago for the study of the
Bose condensation in superfluid $^4$He, attained a considerable
experimental development. The technique proved to be remarkably
adequate to explore the momentum distributions of the nuclei in
condensed matter \cite{Evans,Timms}, and its development continues 
nowadays \cite{Vesuvio,Fielding}. Being it a technique based on
large momentum and energy transfers, the interpretation of the recorded
spectra is based on the validity of the Impulse Approximation (IA),
where it is assumed that the target particle recoils freely after the
collision with the incident particle \cite{Sears71}, assuming that the
binding energy of the target nuclei is negligible compared with the
energy transferred by the neutron to the target nuclei
\cite{Hohenberg,Sears71,Sears73}. The justification of the validity of
the IA in DINS experiments can be found in Refs.  \onlinecite{Platz}
and \onlinecite{Newton}, while the deviations either due to
final-state effects (FSE)
\cite{Sears69,Woods77,Weinstein,Sears84,Mayers89}, or to initial-state
effects (ISE) \cite{Mayers90} were extensively treated.  The IA allows
a direct relation of the scattering law with the momentum distribution
function, making use of the variable $y$, which is the projection of
the initial impulse of the target nuclei over the direction of the
momentum transfer vector $\bm {q}$, through a process known as {\it
  y-scaling} \cite{Sears84,Mayers90}.

On the other hand, the reactor based techniques for the measurement of
atomic momentum distributions (developed before DINS) have employed
triple-axis crystal spectrometers, magnetically pulsed beams, or
rotating crystal and chopper time-of-flight spectrometers \cite{Holt}.
Since these techniques employ neutrons with incident energies
restricted to the thermal range, they do not satisfy the IA and
therefore sophisticated corrections by FSE become essential before the
momentum distribution can be obtained.  The advent of high epithermal
neutron fluxes available from intense pulsed spallation sources,
placed the DINS technique as the most suitable one for the study of
momentum distributions of nuclei constituting the condensed matter.

A common practice in the data treatment of the DINS experiments is to
employ a convolution approximation (CA)
\cite{RAL,Fielding,MonteCarloEllos}, where it is assumed that the
observed spectrum in the time-of-flight scale can be expressed as a
convolution in the variable $y$ of the momentum distribution, with a
mass-dependent {\it ad hoc} resolution function devised to take into
account the experimental uncertainties along with the resonant filter
energy width.  However, the accuracy of this procedure has been
questioned by us in previous papers \cite{Blos1,Blos2}. The criticisms
rested on the fact that the expression of the CA is not directly
deducible from the exact one.  When we tested it against the exact
expression we observed inaccurate results for the peak positions,
areas and shapes of the observed profiles, and the effects were more
noticeable when treating light nuclei. As a consequence, a noteworthy
result of DINS reported on H$_2$O/D$_2$O mixtures \cite{PRLdeellos},
that reported to have found an anomalous behavior in the neutron cross
sections, was critically analyzed from the point of view of the
accuracy of the CA.  A careful analysis lead us to conclude
\cite{Blos1,Blos2} that the inaccuracies committed in the CA produce
anomalies in the same sense as reported. Later measurements performed
by transmission on H$_2$O/D$_2$O mixtures \cite{Letter}, showed no
traces of anomalous neutron cross sections.  Furthermore, on the
theoretical side, recent publications cast doubts on the existence of
anomalous phenomena \cite{Cowley, Colognesi}.  The controversial
situation increases the need to revise the procedures on which the CA
is based.

Besides the mentioned limitations, there is another important
unsatisfactory feature of the CA, that will be treated in the present
paper: in the CA only a single final energy corresponding to the peak
of the resonant filter is considered when the transformation from
time-of-flight to the $y$ scale is performed.  However, as was
remarked in a recent analysis \cite{FinalE}, the final energies of the
scattered neutrons are not restricted to a narrow resonant-filter line
width, but to a broader energy distribution determined by the
kinematics, the dynamics and the full total cross section of the
filter. The contributions of detected neutrons with final energies
outside the main resonance peak cannot be neglected and in some cases
become dominant.

The motivation of the this paper is to expose the details of the exact
treatment that must be performed on the DINS experimental data, in
view of the inconveniences that stem from the mentioned weakness of
the CA.  We reformulate the basic equations that describe the neutron
Compton profiles starting from the exact expressions as was previously
treated \cite{Blos1}. As a result, explicit integrals in the variable
$y$ are obtained.  The passage from the time-of-flight variable $t$ to
$y$ is presented in its exact form, and the final expression results
in a Volterra equation of the first kind, which can be solved for the
momentum distribution $J(y)$.  The results of the present formalism,
which is essentially exact are compared with those produced by the CA.
We stress on the imperfections that it generates and how those
problems are solved by employing the present formalism. The
expressions we present are amenable for the use of the
experimentalists, and we recommend them to replace the currently
employed formalism in the data analysis of DINS experiments.

\section{Theoretical background}
 
\subsection{Preliminary considerations}

In this section we will present a brief summary of the basic equations
that will be employed throughout this paper for describing the
observed spectra in DINS experiments. For a more complete theoretical
development, the reader is referred to Ref.  \onlinecite{Blos1}.  We
will restrict our analysis to a standard inverse-geometry DINS
experimental setup \cite{RAL}, in which incident neutrons with energy
$E_0$ (characterized by the spectrum $\Phi(E_0)$) travel along a
distance $L_0$ from the pulsed source to the sample.  Scattered
neutrons at an angle $\theta$ and a final energy $E$ travel along a
distance $L_1$ up to the detector position. A movable filter with a
neutron absorption resonance in the eV-energy region is placed in the
scattered neutron path, so consecutive `filter in' and `filter out'
measurements are performed. The spectra are recorded as a function of
the total time of flight $t$. Throughout this paper, we will
explicitly omit the description of experimental uncertainties due to
geometry, time of flight or multiple scattering. For an account on
these effects the reader is referred to Refs.
\onlinecite{MonteCarloEllos} and \onlinecite{MonteCarloNuestro}.

The measured magnitude (known as Neutron Compton Profile (NCP)) is the
difference count rate (`filter out' minus `filter in') as a function
of $t$, which for a point-like sample can be expressed as \cite{Blos1}
\begin{widetext}
\begin{equation}
\label{ct}
c(t)=\frac {\sqrt{8/m}}{L_1}\Delta\Omega\, \mathop{\int_{\scriptstyle E_{\rm 0\,inf}}^{+\infty}}_{\rm t=const}
dE_0\,E^{3/2} \,\Phi(E_0) \,\sigma(E_0,E,\theta) \,\varepsilon(E) \left[1-e^{-nT\sigma_F(E)} \right]
\end{equation}
\end{widetext}
where $\sigma(E_0,E,\theta)$ is the sample double-differential cross section,
$\varepsilon(E)$ the detector efficiency and $\Delta\Omega$ the solid angle subtended
by the detector. The term between brackets is the absorption
probability of the resonant filter, characterized by a number density $n$,
a thickness $T$  and a total cross section $\sigma_F(E)$. As indicated,
the integral in Eq. (\ref{ct}) must be calculated at a constant
time-of-flight $t$ given by the kinematic condition
\begin{equation}
\label{tof}
t=\sqrt{\frac{m}{2}}\left(\frac{L_0}{\sqrt{E_0}}+\frac{L_1}{\sqrt{E}} \right),
\end{equation}
where $m$ is the neutron mass.
The lower limit of integration $ E_{\rm 0\,inf}$ is determined from the
kinematic condition that in the second flight path the neutron has
infinite velocity. Its value is
\begin{equation}
\label{E0inf}
E_{\rm 0\,inf}=\frac{m}{2}\left(\frac{L_0}{t}\right)^2.
\end{equation}

As usual, let
\begin{equation}
\label{hw}
\hbar\omega=E_0-E
\end{equation}
be the energy transferred by the neutron to the sample, and
\begin{equation}
\label{hq}
\hbar q=\sqrt{2m\left(E_0+E-2\sqrt{E_{0}E}cos\theta\right)}
\end{equation}
the modulus of the transferred impulse. The DINS technique was
developed on the theoretical basis of the IA, in which the scattering
law (for a monatomic sample) can be written \cite{Mayers90}
\begin{equation}
\label{Sqw4}
S(\mathbf{q},\omega)=\frac{M}{\hbar q}  J\left(\mathbf{e_q},y\right),
\end{equation}
where $\bm{e_q}$ is the unit vector along the direction of $\bm{q}$.
The variable $y$ is the projection of the impulse $\bm{p}$ of the nucleus
of mass $M$ on $\bm{e_q}$, and can be expressed as
\begin{equation}
\label{y}
y=\frac{M}{\hbar q}  \left(\hbar\omega-\frac{\hbar^2q^2}{2M}\right).
\end{equation}
$J(\bm{e_q},y)dy$ is defined as the probability that a nucleus has a
momentum component along the direction of $\bm{e_q}$ with values
between $y$ and $y+dy$.  The distribution $J\left(\bm{e_q},y\right)$
must be symmetric around $y=0$ if a moment in a given direction of
the space is equally probable than in its opposite. For an isotropic
sample $J\left(\bm{e_q},y\right)$ does not depend on the direction
given by $\bm{e_q}$, and then $J(\bm{e_q},y)dy=J(y)dy$ is the
probability that a nucleus of the sample has component of momentum
between $y$ and $y+dy$ along any direction in the space. So, the
dynamic structure factor is reduced to
\begin{equation}
\label{Sqw5}
S(\mathbf{q},\omega)=\frac{M}{\hbar q}  J\left(y\right).
\end{equation}
Then, for an isotropic sample composed of different nuclei, being $N_M$
the number of nuclei of mass $M$ and bound scattering length $b_M$,
the double differential cross section is
\begin{equation}
\label{Sigma}
\sigma(E_0,E,\theta)=\frac{1}{\hbar q} \sqrt{\frac{E}{E_0}} \sum_{M}{N_M}{b_M^2}M J_M\left(y_M\right),
\end{equation}
where the sum is extended over all the masses corresponding to the
different species (in non equivalent positions) present in the sample.

\subsection{Convolution approximation}
\label{CA}

The usually employed convolution approximation is expressed by \cite{RAL}

\begin{equation}
\label{convo}
c_{conv}(t)= \sum_M \xi_M(t) J_M(\tilde y_M) \otimes R_M(\tilde y_M),
\end{equation}
with
\begin{equation}
  \label{xi}
  \xi_M(t)=N_Mb_M^2 M \left[\Phi(E_0)\frac{dE_0}{dt}\varepsilon(E_1)\Delta\Omega \,\Delta E_1 \sqrt{\frac{E_1}{E_0}}\frac{1}{q}\right],
\end{equation}
where $E_1$ is fixed, defined by the energy of the main absorption
peak in the filter total cross section. The sum has the same meaning
as in Eq.  (\ref{Sigma}), and a resolution function $R(\tilde y_M)$ is
introduced as a way to contain the geometric uncertainties as well as
the filter resonance width. It is worth to emphasize that that in Eq.
(\ref{convo}), the relation between the variable $\tilde y_M$ and $t$
is made through Eq. (\ref{y}), where $q$ and $\omega$ are calculated also
with the same $E_1$ fixed, and $E_0$ compatible with the
time-of-flight relation (\ref{tof}). Also, it must be noticed that the
term between square brackets in Eq. (\ref{xi}) is independent of the
sample characteristics.

It is important to notice that the resolution function in the CA
framework is deducible from Eq. (\ref{convo}) and Eq. (\ref{ct})
considering a sample represented by an ideal gas in the limit of $T\to$
0K. In this case $J_M(\tilde y_M) \to \delta(\tilde y_M)$ and \cite{Blos1}
\begin{equation}
  \label{fres}
  R_M(\tilde y_M)=\frac{c(t)_{T=0K}}{\xi_M(t)}.
\end{equation}
When geometric uncertainties are considered, they also contribute to
the width of $R_M$, but as we have already mentioned, this case will
not be analyzed in the present paper. It is worth to comment that in
common practice Eq. (\ref{fres}) is not employed, but instead a fitted
Lorentzian (for Gold filters) or Gaussian function (for Uranium) in
the variable $y$ are employed \cite{Timms,Fielding}.

\section{Formulation of the exact expression}

As it was commented, Eq. (\ref{convo}) can not be deduced from the
exact expression (\ref{ct}) and therefore its application can not be
fully justified. On the other hand an expression where (like in
(\ref{convo})) the $y$ variable appears explicitly, is desirable since
it is the natural variable of the impulse distribution $J(y)$. In this
section we will deduce such exact expression that keeps $y$ as the
integration variable. To this end, we will change the integration
variable from $E_0$ to $y$ in Eq.  (\ref{ct}).  In the first step we
rewrite Eq. (\ref{y}) in terms of $E$, $E_0$ and $\theta$ as

\begin{equation}
\label{ydeE0}
y_M=\frac{M\left(E_0-E-\displaystyle \frac{m}{M}\left(E_0+E-2\sqrt{{E_0}E}\cos\theta\right)\right)}
{\sqrt{2m\left(E_0+E-2\sqrt{{E_0}E}\cos\theta\right)}}.
\end{equation}

At a given time of flight $t$, $E_0$ and $E$ are linked through Eq.
(\ref{tof}). Therefore, at a scattering angle $\theta$, $y_M$ is a
function of $E_0$ and $t$. On the other hand it must be noticed that
$E_0$ can be a multi-valuated function of $y_M$. There is a
substantial difference between this definition of $y_M$ and that
employed in the CA (Sect. \ref{CA}), called $\tilde y_M$. In the
present definition $y_M$ is a function of $E_0$ for each considered
time of flight, while in the former definition $\tilde y_M$ takes one
value for each $t$.

To proceed with the variable change, we first study the limits of
integration in the variable $y$.  When $E_0\to E_{\rm 0\,inf}$ (defined in
Eq.(\ref{E0inf})), $y\to -\infty$ regardless the mass of the scattering
nuclei. When $E_0\to+\infty$, it can be readily
shown that

\begin{equation}
\label{ylimites}
\mathop{\lim_{E_0\to+\infty}}_{t,\theta=\rm{const.}}{y_M}=\left\{
\begin{array}{ll}
+\infty & {\rm if\ } M>m;\\
-\infty & {\rm if\ } M<m;\\
m\frac{L_1}{t}\cos\theta & {\rm if\ } M=m.
\end{array}
\right.
\end{equation}
In Sect. \ref{ymap} we will present a detailed study of the behavior
of $y_M$.

The sought variable change in Eq. (\ref{ct}) involves the Jacobian
(calculated in Appendix \ref{App})

\begin{widetext}
\begin{equation}
\label{dyde01}
\displaystyle \frac{dy}{dE_0}\Bigg|_{\rm{t},\theta}=
\frac{-m}{\hbar q} \left(\frac{M\hbar\omega}{{\hbar^2}q^2}+\frac{1}{2}\right)
\left[1-\frac{L_0}{L_1}\left(\frac{E}{E_0}\right)^{3/2}+\left(\frac{L_0}{L_1}\frac{E}{E_0}-
\sqrt{\frac{E}{E_0}}\right)\cos\theta\right]+\frac{M}{\hbar q}\left[1+\frac{L_0}{L_1}
\left(\frac{E}{E_0}\right)^{3/2}\right].
\end{equation}
\end{widetext}
To perform the variable change from $E_0$ to $y$ in the integral of
Eq. (\ref{ct}), it is necessary to know the intervals where the
function $y(E_0)$ is monotonous. Therefore we must obtain the $E_0$
values (and its corresponding $y$ values) where 
\begin{equation}
  \label{roots}
  \displaystyle\frac{dy}{dE_0}\Bigg|_{\rm{t},\theta}=0.  
\end{equation}

Defining the variables
\begin{equation}
\label{adimen}
\left\{
\begin{array}{lll} 
x&=&\sqrt{\frac{\displaystyle E}{\displaystyle E_0}} \nonumber \\
\ell&=&\frac{\displaystyle L_0}{\displaystyle L_1} \nonumber \\
\end{array}
\right.,
\end{equation}
(where the positive value of the square root is taken in the
definition of $x$), Eq. (\ref{roots}) can be reduced to a fifth-order
polynomial in $x$, with real coefficients. The sought values of $E_0$
are obtained from the roots of
\begin{equation}
\label{poli1}
\mathop{\sum}_{i=0}^{5}a_i{x^i}=0
\end{equation}
where
\begin{eqnarray}
\label{coef1}
a_0&=&m-M  \nonumber \\
a_1&=&3(M-m)\cos\theta \nonumber \\
a_2&=&m-3M+(m+M)\ell\cos\theta+2m\cos^2\theta  \\
a_3&=&(M-m)\cos\theta-(3M+m)\ell -2m\ell \cos^2\theta \nonumber  \\
a_4&=&3\ell(m+M)\cos\theta \nonumber \\
a_5&=&-\ell (m+M).  \nonumber
\end{eqnarray}
Therefore, Eq. (\ref{poli1}) has at most five real roots and at least
one, and only the positive values have physical sense.  Let $N_R$ be
the number of such roots ($0\leq N_R\leq5$), whose values will be labeled
$x_j$, and correspondingly $E_0^j$ and $y_{M,j}$ the roots in the
$E_0$- and $y$-scales respectively.  Now we can proceed to change the
integration variable from $E_0$ to $y$ in Eq. (\ref{ct}). To this end,
the integration range has to be partitioned in $N_R+1$ intervals
$(E_0^{j-1},E_0^j)$, where the function $y_M(E_0)$ is monotonous, which
we will identify as `branch $j$' (where $E_0^0=E_{\rm 0\,inf}$ given
in Eq.(\ref{E0inf}) and $E_0^{N_R+1}=+\infty$). The result is

\begin{widetext}
\begin{equation}
\label{cdety1}
c(t)=\frac{\sqrt{8/m}}{L_1}\Delta\Omega\sum_{M}\sum_{j=1}^{N_R+1}
\mathop{\int_{\displaystyle y_{M,j-1}}^{\displaystyle y_{M,j}}}_{\rm t=const}dy_M\frac{dE_0}{dy_M}\Bigg|_{t}^{j}
 \,\Phi(E_0)\sqrt{\frac{E}{E_0}} \,\left[1-e^{-nT\sigma_F(E)} \right] \,\varepsilon(E) E^{3/2}
 \frac{1}{\hbar q} N_{M} b_{M}^{2}MJ_{M}(y_M),
\end{equation}
where the superscript $j$ in $dE_0/dy_M$ indicates the branch where it
must be evaluated, $y_{M,0}=-\infty$ and $y_{M,N_R+1}$ is defined in Eq.
(\ref{ylimites}). Thus, the NCP results

\begin{equation}
\label{cdety2}
c(t)=\sum_{M}\sum_{j=1}^{N_R+1}
\mathop{\int_{\displaystyle  y_{M,j-1}}^{\displaystyle y_{M,j}}}_{\rm t=const}dy_M f_M^{j}(y_M,t)J_{M}(y_M)
\end{equation}
where the function
\begin{equation}
\label{fMj}
f_M^{j}(y_M,t)=\frac{\sqrt{8/m}}{L_1}\Delta\Omega \frac{1}{\hbar q} N_{M} b_{M}^{2}M\frac{dE_0}{dy_M}\Bigg|_{t,\theta}^{j}
 \,\Phi(E_0)\sqrt{\frac{E}{E_0}} \,\left[1-e^{-nT\sigma_F(E)} \right] \,\varepsilon(E) E^{3/2}
\end{equation}
\end{widetext}
is defined in the interval $(y_{M,j-1},y_{M,j})$, and is zero outside
such range.

Finally, for a sample containing $N_M$ atoms of mass $M$ in non
equivalent positions, the NCP can be expressed as

\begin{equation}
\label{cdety3}
c(t)=\mathop{\sum}_{M}c_M(t)
\end{equation}
where
\begin{equation}
\label{cdety3}
c_M(t)=\mathop{\int_{\scriptstyle {-\infty}}^{+\infty}}_{\rm t=const}dy_M f_M(y_M,t)J_{M}(y_M)
\end{equation}
and $f_M(y_M,t)$ is the sum on the different branches ($1\leq j \leq
N_R+1$) of $f_M^{j}(y_M,t)$.

Eqs. (\ref{cdety2}) and (\ref{fMj}) are the central expressions of the
present paper, which are exact expressions, since they were derived
directly from Eq. (\ref{ct}) without any approximation. Eq.
(\ref{cdety2}) is a Volterra equation of the first kind. To obtain
$J(y)$ from it, it is possible to resort to numerical methods.
Alternatively, if the function that describes $J(y)$ is known
beforehand (or a plausible form is assumed for it), it will be
possible to obtain it from a fitting procedure. In this last case it
will not be necessary to change the integration variable to $y$. In
the case where the sample and/or detector dimensions are not
negligible, Eqs. (\ref{cdety1}-\ref{cdety3}) must be integrated also
over the geometric dimensions. 

As it was already mentioned, the definition of $y_M$ employed in
Eqs. (\ref{cdety1}-\ref{cdety3}) differs from that employed in the
CA. It is worth to notice that while in the CA the resolution function
is the same for every $t$ (centered in $\tilde y_M(t)$), in the
present formalism, $f_M(y_M,t)$ depends explicitly on $t$.
It must be remarked that if the CA framework were exact, Eqs.
(\ref{convo}) and (\ref{cdety2}) would produce the same NCP, for an
arbitrary $J(y)$. Thus, from a direct comparison between Eqs.
(\ref{cdety3}) and (\ref{convo}) the following relation between the
exact expression of $f_M(y_M,t)$ and the resolution function would be
verified
\begin{equation}
  \label{convvsex}
  \xi(t) R_M(\tilde y_M(t)-y)=f_M(y,t)
\end{equation}
where $\xi_M(t)$ is a defined in Eq. (\ref{xi}). However, the usually
employed $R_M(y_M)$ does not verify expression (\ref{convvsex}), as
will be examined in the next section.

\section{Discussion}

In this Section we will analyze different aspects of the formalism we
are presenting, with especial emphasis on the differences between the
exact results and those obtained in the CA framework. We will focus
our attention on scattering on H, D and $^3$He, since as we will show,
they are the most sensitive cases in normal experimental conditions.
We will assume throughout detectors of ideal efficiency $\varepsilon(E)=1$. An
incident neutron spectrum described by $\Phi(E_0)=E_0^{-0.9}$, and flight
lengths $L_0$ = 1105.5 cm and $L_1$ = 69 cm (so $\ell=16.022$) will be
chosen to match the layout of the DINS facility at the Rutherford
Appleton Laboratory \cite{Fielding}. The filter will be represented
with a gold foil of $\rm nT=4\times 10^{-5}$ barn$^{-1}$ and its total
cross section will be described from the data compiled in Ref.
\onlinecite{ENDF}.  

We will divide our discussion in three different aspects of our
formalism. The first will be devoted to a close examination of the
behavior of the variable $y$. We will establish the relationship
between $y$ in the exact formalism and in the CA framework. In the
second part we will analyze the behavior of the kernel $f_M(y,t)$ and
we will compare it with the resolution function employed in the CA.
In the third part we will analyze the consequence of our analysis in a
specific case.  The starting result over which we will base the
subsequent discussions are the NCPs of hydrogen, deuterium (Sects.
\ref{ymap} and \ref{kernf}) and $^3$He (Sect. \ref{meankin}) as
calculated from Eq. (\ref{ct}).

\begin{figure}
\resizebox{0.5\textwidth}{!}{\includegraphics{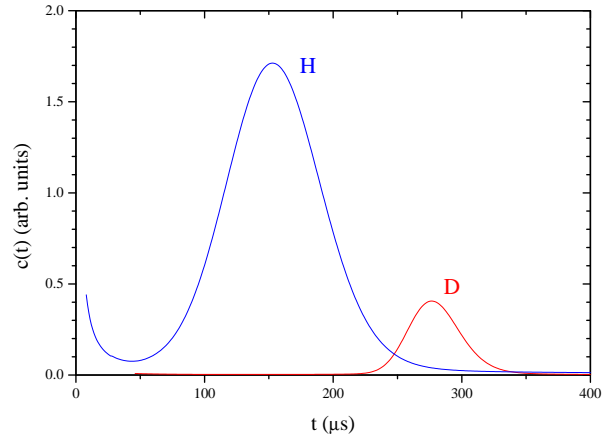}}
\caption{(Color online) Neutron Compton profiles of hydrogen and deuterium at a scattering
  angle of $\theta=70^{o}$ calculated considering as flight paths $L_0$ =
  1105.5 cm and $L_1$ = 69 cm, assuming a gold foil analyzer of $\rm
  nT=4\times 10^{-5}$ barn$^{-1}$ and effective temperatures of 115.22 and 80.11 meV
  respectively.}
\label{figncp}
\end{figure}

The study of H and D will be based on ideal gases of effective
temperatures of 115.22 and 80.11 meV respectively, as described in
Ref.  [\onlinecite{Blos1}]. Their NCPs are shown in Fig. \ref{figncp}.

\subsection{Study of the variable $y_M$}
\label{ymap}

\begin{figure}

\resizebox{0.5\textwidth}{!}{\includegraphics{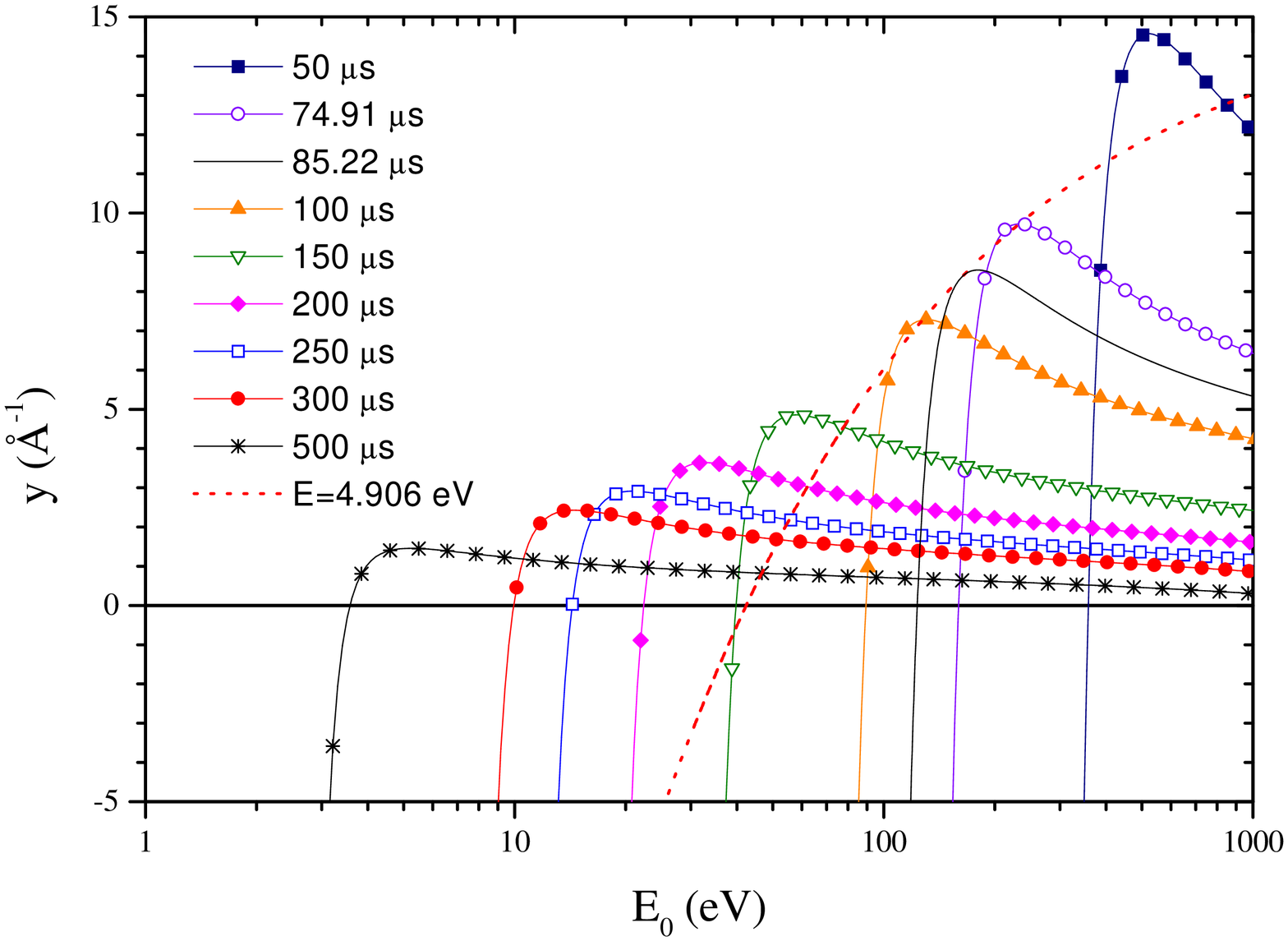}}
\caption{(Color online) The impulse $y$ as a function of the incident energy $E_0$ for A=0.99862. The       
  exact calculations according to Eq. (\ref{ydeE0}) at the indicated
  times of flight are compared with the calculation performed by
  assuming the final energy fixed at 4.906 eV, which is the variable
  $\tilde y$ employed in the CA framework}
\label{figye0}
\end{figure}

We will begin our discussion by examining the behavior of the $y_M$
variable as a function of the time of flight. In Fig. \ref{figye0} we
show its behavior according to Eq. (\ref{ydeE0}) as a function of the
incident neutron energy $E_0$ at different times of flight, for the
case of hydrogen (M/m=0.99862) at $\theta=$70{\r{}}.  The behavior at the
limits commented in Eq.(\ref{ylimites}) are observed.  In the same
figure we show the behavior of $\tilde y_M$ according to the
conditions normally employed in the CA framework (Sect.  \ref{CA}).
The differences between both ways to define $y$ is evident: while in
the exact formalism there is a whole range of possible values of $y$
at each time of flight (which is a function of $E_0$), in the CA there
is only one value of $y$ defined at each $t$.  We observe the curves
$y(E_0)$ in the exact formalism intersects the one of the CA in two
points.  However at each time of flight there is only one value of $y$
defined in the CA framework. Thus, only one of those two points over
the exact $y$-curve has a final energy corresponding to that of the
filter (4.906 eV), while the other has the same value of $y$ and $E_0$
but a different $E$. At 85.22 $\mu$sec, both points coincide so the
curves are tangent.  At times of flight greater than 85.22 $\mu$sec,
this property is found in the intersection of the leftmost point
between both curves, while at times lower than 85.22 $\mu$sec we find
it in the rightmost intersection point.

\begin{figure}
\resizebox{0.5\textwidth}{!}{\includegraphics{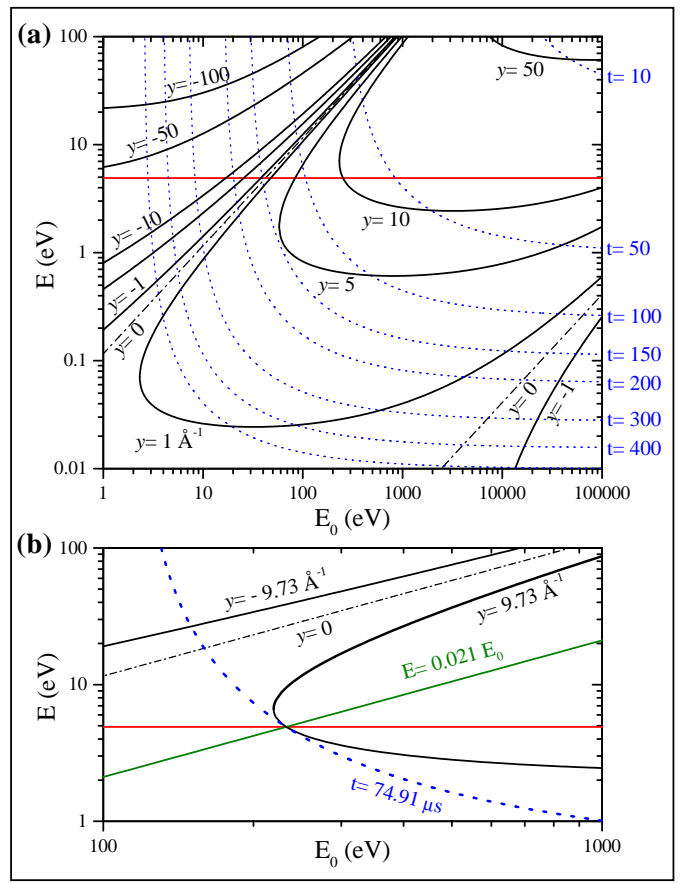}}
\caption{(Color online) (a) The full thick lines indicate the constant-$y$ curves
  for hydrogen at $\theta =70^{o}$ and $\ell=16.022$. $y$-values are indicated
  in {\AA}$^{-1}$.  In dotted lines, the curves of $t=$ constant (the
  indicated values are in $\mu$sec).  The horizontal full line corresponds
  to E=4.906 eV. (b) Detail of the upper frame showing a case where a
  constant-$t$ and a constant-$y$ curve are tangent. The particular
  case of a tangency at $E$=4.906 eV was selected (see text for
  details). }
\label{figmap1}
\end{figure}

The behaviors observed in Fig. \ref{figye0} are clarified in Fig.
\ref{figmap1}(a), where we mapped on the $(E_0,E)$-plane the
constant-$y$ and constant-$t$ curves, for the case of hydrogen. In
Appendix \ref{App2} we show the details for the calculation of the
constant $y$ curves.  We also show the constant-$E$ line corresponding
to main peak absorption energy of the gold analyzer (E=4.906 eV). The
intersection of a constant-$t$ curve with the 4.906 eV-line defines an
$E_0$-value with which the $y$-value in the CA framework is
calculated. It can be easily shown that in this case for each
constant-$t$ curve, there is one (and only one) tangent constant-$y$
curve. The geometric loci of the tangency points in the
$(E_0,E)$-plane result to be straight lines (which will be called {\it
  limit lines} (LL) hereafter) whose slopes can be obtained from the
real positive roots of Eq.  (\ref{poli1}).  Bearing in mind Eq.
(\ref{adimen}) those slopes are the squares of such roots. In this
particular case of hydrogen $\displaystyle
\frac{dy}{dE_0}\Bigg|_{\rm{t},\theta}=0$ has only one real root, which
defines the line $E=0.021E_0$ as the sought geometric locus of the
tangency points, which is represented in Fig.  \ref{figmap1}(b).  In
the representation of Fig.  \ref{figye0} this corresponds to the
geometric locus of the maxima of the $y(E_0)$ curves. It is worth to
point out that the consequence of having one real root in Eq.
(\ref{poli1}) is that there are two branches, so $N_R=1$ in Eq.
(\ref{cdety2}).  In Fig.  \ref{figmap1}(b) we show a detail of the
upper frame. The line $E=0.021E_0$ intersects the constant $E=4.906
eV$ at a point corresponding to $t=$74.91 $\mu$sec and $y=$ 9.73
{\AA}$^{-1}$.  In consequence, in Fig.  \ref{figye0} the intersection
between $\tilde y$ and the curve of 74.91 $\mu$sec occurs at its
maximum.  The importance of the LL resides in that it indicates the
$y$ values where $f_M(y,t)$ presents singularities, and will be
commented in the next section.

\begin{figure}
\resizebox{0.5\textwidth}{!}{\includegraphics{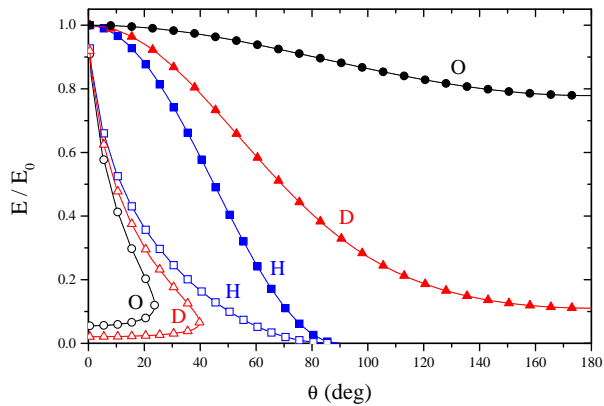}}
\caption{(Color online) The open symbols indicate value of the slopes of the straight
  lines corresponding to the geometric loci of the tangency points
  between constant-$t$ and constant-$y$ curves in the $(E,E_0)$-plane
  as a function of the scattering angle, for H, D and O, with a
  flight-paths ratio $\ell=16.022$. Full symbols indicate the ratio
  $E/E_0$ in the cold-gas limit for H, D and O.}
\label{figeth}
\end{figure}

The slope of the LLs is a function of the scattering angle and $\ell$.
This is shown in Fig. \ref{figeth}, for $\ell=16.022$. In the case of
hydrogen a LL can be defined in the whole angular range from 0{\r{}} to
90{\r{}} existing only one of such LLs as already mentioned. On the other
hand, for deuterium there exist two real positive solutions of Eq.
(\ref{poli1}) below 39.7{\r{}} and so two LLs, and for oxygen we observe
the same situation at a limit angle of 24.7{\r{}}.

\begin{figure}
\resizebox{0.5\textwidth}{!}{\includegraphics{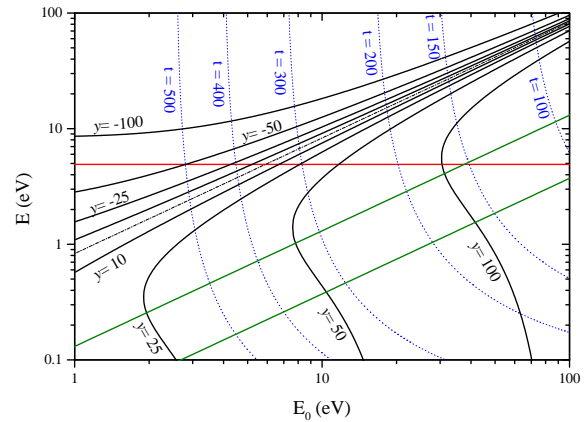}}
\caption{(Color online) The full thick lines indicate the constant-$y$ curves
  for deuterium at $\theta =35^{o}$ and $\ell=16.022$. $y$-values are
  indicated in {\AA}$^{-1}$.  In dotted lines, the curves of $t=$ constant
  (the indicated values are in $\mu$sec). The line corresponding to
  $y=$0 is shown in dash-dotted line.}
\label{figmap2}
\end{figure}

It is also interesting to observe in Fig. \ref{figmap2} the
representation in the $(E_0,E)$-plane for deuterium at $\theta=$35{\r{}}. In
this case there exist two LLs (in accordance with Fig. \ref{figeth}),
which are shown in the graph.

\subsection{Kernel $f_M(y_M,t)$}
\label{kernf}

\begin{figure}
\resizebox{0.5\textwidth}{!}{\includegraphics{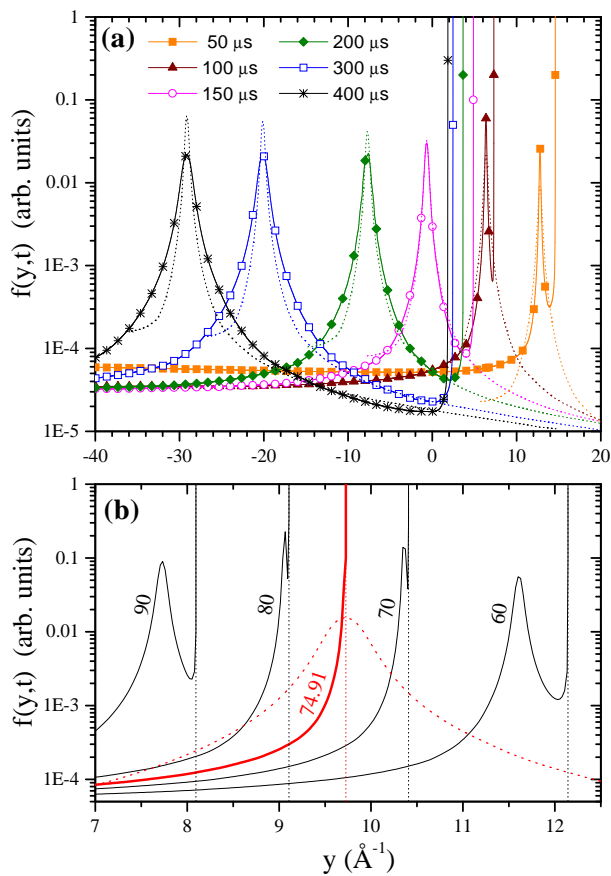}}
\caption{(Color online) (a) Full lines: $f(y,t)$ at $\theta =70^{o}$ for the indicated times of flight.    
  Vertical dotted lines: position of the asymptotes of $f(y,t)$.
  Dotted lines: Resolution function times the amplitude factor
  $\xi(t)$, centered at the $\tilde y_M(t)$ values.  See text for
  details. (b) Detail of the upper frame. Full thick line: $f(y,t)$
  for the special case of t=74.91 $\mu$s, for which the asymptote
  position agrees with the value of $y$ where the resolution function
  times $\xi(t)$ (dotted thick line) is centered. Full thin lines:
  $f(y,t)$ where the values of $t$ are indicated in $\mu$sec. }
\label{figfyt}
\end{figure}

\begin{figure}
\resizebox{0.5\textwidth}{!}{\includegraphics{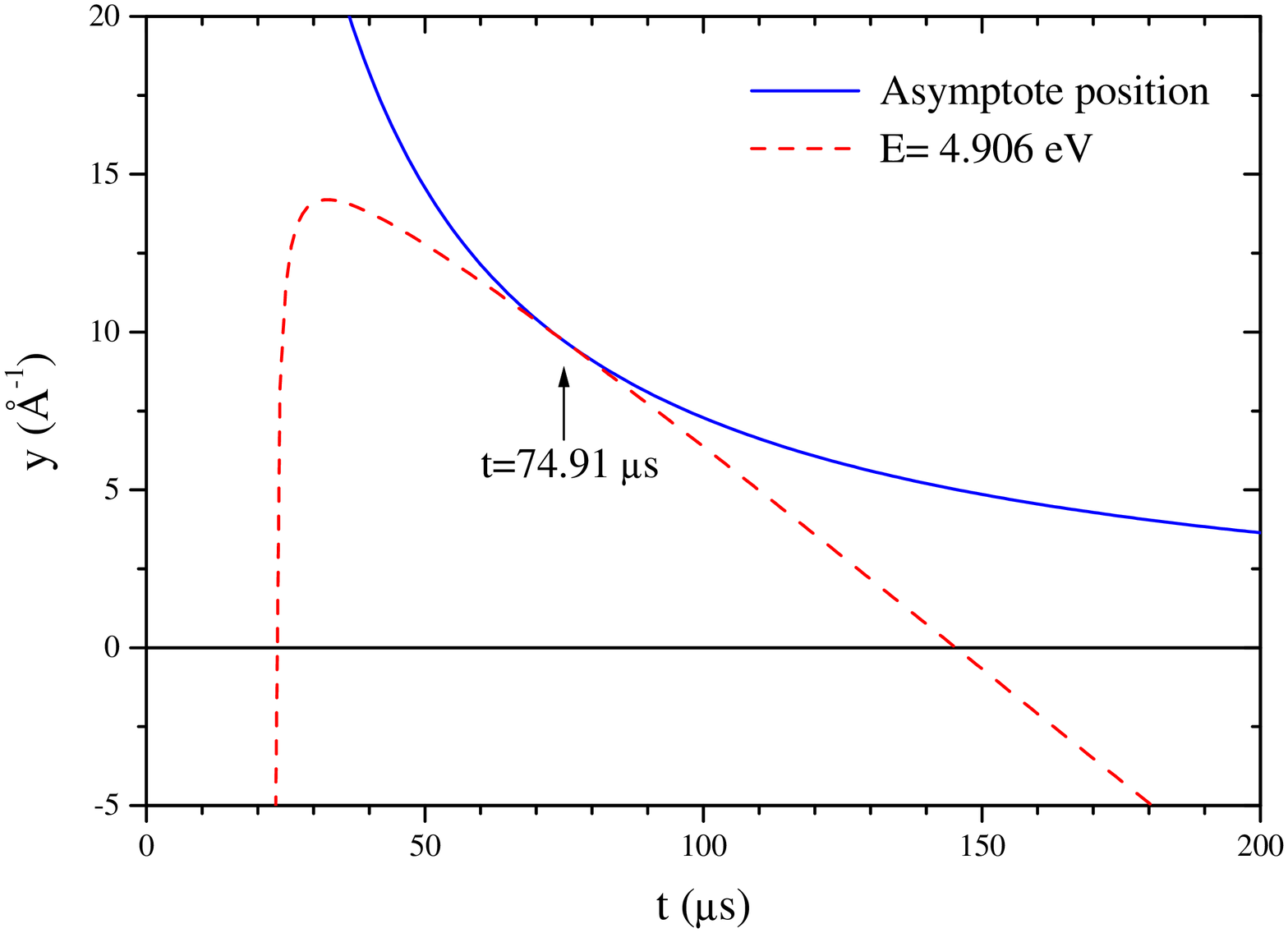}}
\caption{(Color online) Asymptote position of $f(y,t)$ as a function of the time of
  flight $t$, for the case of hydrogen at a scattering angle of
  $\theta=70^{o}$, and flight length of $L_0$ = 1105.5 cm and $L_1$ = 69
  cm.}
\label{figasy}
\end{figure} 

Having established the behavior of the variable $y_M$ we can now
analyze Eq. (\ref{cdety3}). We are especially interested in examining
the kernel $f_M(y_M,t)$. In Fig. \ref{figfyt} we show its behavior in
the case of scattering in hydrogen at 70{\r{}} for different times of
flight of interest in the Compton profile, as can be checked in Fig.
\ref{figncp}. It is very important to notice the asymptotic behavior
of $f_M(y_M,t)$ at the values of $y_M$ defined by the LL, that is
understood from the fact that the Jacobian $\displaystyle
\frac{dE_0}{dy}\Bigg|_{\rm{t},\theta}\to \infty $ in Eq. (\ref{fMj}). In the
same figure we show the corresponding kernel in the CA framework
expressed in the left hand side of Eq.  (\ref{convvsex}), where the
resolution function was calculated according to Eq(\ref{fres}). The
maxima of both distributions are observed to be located at $\tilde
y_M(t)$, {\it i.e.} the $y$ variable as defined in the CA framework.
However, the asymptotes present in the exact formulation mark
important differences between both formulations, which are clearly
manifested at 74.91 $\mu$sec, as showed in Fig.  \ref{figfyt}(b), where
the asymptote is exactly in $\tilde y_M(t)$.  This effect is further
illustrated in Fig. \ref{figasy}, where the asymptote position is
shown as a function of $t$ together with $\tilde y_M(t)$. Both curves
are tangent at 74.91 $\mu$sec. Alternatively, the interpretation of
this point can be understood from the inspection of Fig.
\ref{figmap1}(b). In the $(E_0,E)$-plane this point is the
intersection between the LL and the $E=$4.906-eV line. On any
constant-$t$ curve the maximum value of $y$ occurs in its intersection
with the LL (point A), so in general this value of $y$ is greater than
the one at its intersection with the constant line $E=$4.906 eV (point
B). But at the precise time where both lines intersect points A and B
coincide.  This is manifested in Fig. \ref{figasy} where the $y$
values of the asymptote position are always greater than those of the
constant line $E=$4.906 eV except at the tangency point.

Besides the mentioned asymptotic behavior, there exist two other
important features not described in the CA framework, {\it viz.}  the
main peak width and height depends on $t$, at variance with the left
hand side member of Eq. (\ref{convvsex}). Similarly, those behaviors
are observed in deuterium (not shown in this paper).

\begin{figure}
\resizebox{0.5\textwidth}{!}{\includegraphics{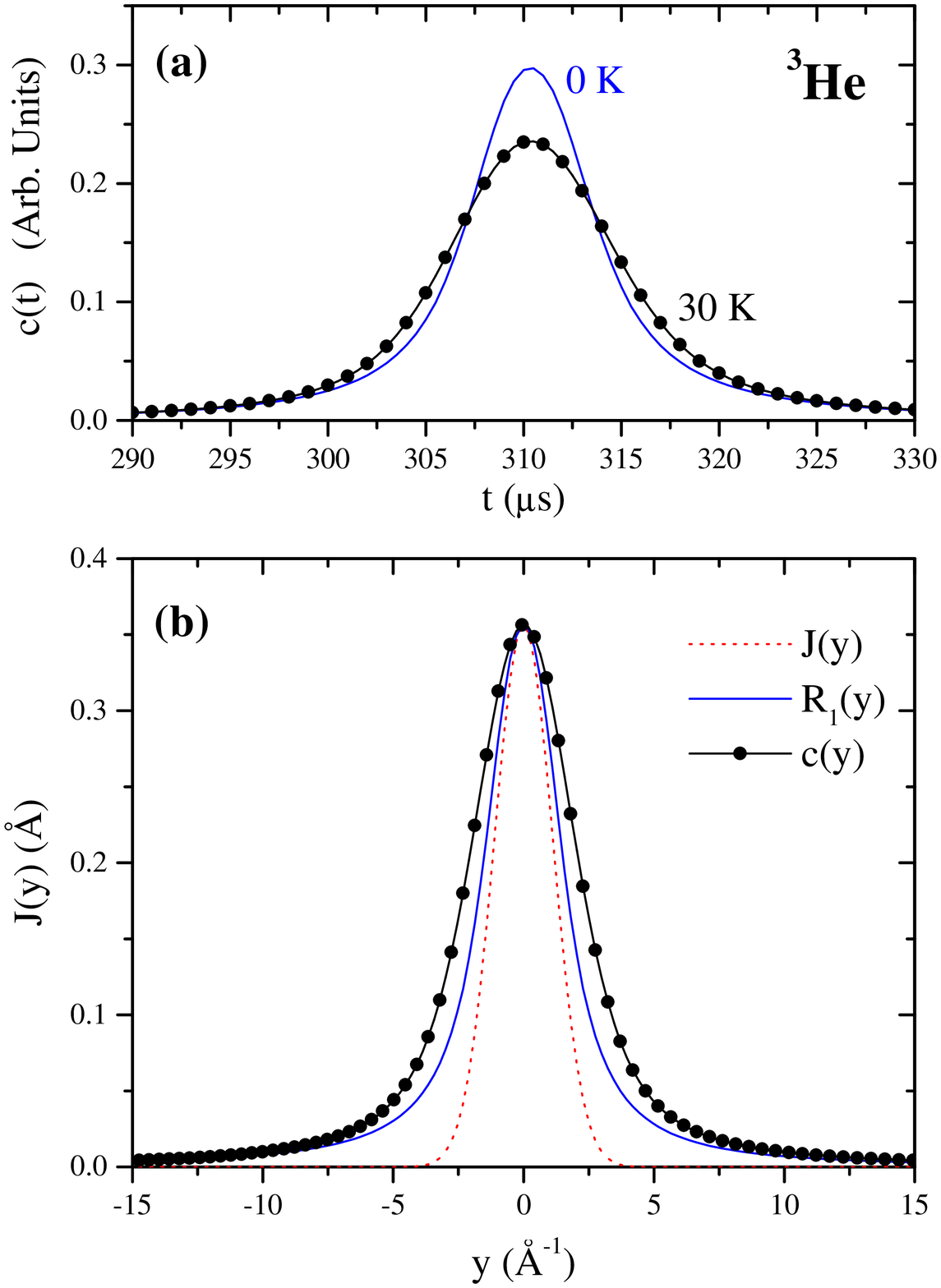}}
\caption{(Color online) (a) Neutron Compton profiles in the time of
  flight scale of $^3$He at a scattering angle of 70{\r{}}, at 0K (line)
  and 30 K (circles). (b) The dotted line indicates the Gaussian
  momentum distribution $J(y)$ of $^3$He with a mean kinetic energy of
  30 K.  In full line, the resolution function calculated according to
  Eq.(\ref{fres}) and $c(t)_{T=0 K}$ shown in frame (a).  Circles
  indicate the NCP calculated at 30 K expressed as a function of the
  $y$ variable in the CA framework.}
\label{figres}
\end{figure} 

\subsection{Study of the mean kinetic energy}
\label{meankin}

An immediate consequence of the present analysis can be found in the
study of the mean kinetic energies, which is the main aim of the DINS
technique. We will focus our attention to a sample constituted by an
ideal gas of $^3$He with a mean kinetic energy of 30 K ($k_BT=$1.7234
meV).  Such system was extensively examined in the literature
\cite{Sene, Wang, Sokol}.  The NCP corresponding to this system can be
obtained in the exact formalism with Eq. (\ref{cdety2}) and is shown
in Fig. \ref{figres}(a) (circles), which corresponds to a Gaussian
distribution $J(y)$ with a full-width-at-half-maximum (FWHM) of 1.3151
{\AA}$^{-1}$. The geometry, the filter and the efficiency are the same
described at the beginning of this section. The function $J(y)$ is
shown Fig. \ref{figres}(b)(dotted line).

We now compare the exact formalism with the results produced in the CA
framework. The resolution function can be obtained from the limit $T \to$
0K of the NCP (shown in full line in Fig. \ref{figres}(a)), and it is
calculated in Eq. (\ref{fres}), thus obtaining the $R_1(y)$ shown in Fig.
\ref{figres}(b) (full line). The total cross section for gold employed
in its calculation was taken from Ref. \onlinecite {ENDF}. Following
the procedure already described \cite{Blos1}, Eq. (\ref{convo}) was
employed to perform a least-squares fit of a Gaussian $J(y)$  on the
exact NCP, letting the area and the width of the Gaussian as free
parameters. Despite being $R_1$ the only resolution function in the CA
framework compatible with the limit at $T \to$0 K, we tested three
other different resolution functions based on the customary use

\begin{table}[]
  \centering
  \begin{ruledtabular}
  \begin{tabular}[b]{cc}
  CA & Mean Kinetic Energy (meV)\\
  \hline  
  $R_1$ &  1.694 \\
  $R_2$ &  1.374 \\
  $R_3$ &  1.353 \\
  $R_4$ &  1.519 \\ 
  \hline \hline
  Exact value & 1.7234  \\
  \end{tabular}
  \end{ruledtabular}
  \caption{Mean kinetic energy of $^3$He at 30 K obtained in the 
    CA framework according to different resolution functions 
    described in the text. The exact value is includes for comparison.}
  \label{tab:1}
\end{table}

\begin{itemize}
\item $R_2(y)$, obtained by fitting a Lorentzian function in the
  energy scale \cite{PRLdeellos,dd} to the main resonance peak at 4.906 eV of the gold
  total cross section\cite{ENDF}, and employing Eq (\ref{fres}).
\item $R_3(y)$, a Lorentzian function in the $y$ variable
  \cite{Timms,Evans,Fielding} fitted to $R_2(y)$. The resulting FWHM is
  3.292 {\AA}$^{-1}$.
\item $R_4(y)$, a Lorentzian function in the $y$ variable
  \cite{Timms,Evans,Fielding} fitted to $R_1(y)$. The resulting FWHM is
  3.170 {\AA}$^{-1}$.
\end{itemize}

In table \ref{tab:1} we show the resulting effective temperatures,
compared with the exact value, used as input to generate $J(y)$. It is
worth to remark that the different choices of $R(y)$ produces large
variations in the effective temperature, even when the FWHM has no
such large variations. For example, the change in FWHM from $R_3$ to
$R_4$ is 3.7~\% and the resulting mean kinetic energies differ in
12.3~\%. The reason for such a large change, was reported in Ref.
\onlinecite{Blos1}: at low temperatures the shape of the observed NCP
is determined mainly by the filter, so any imperfection in the
description of the filter effects will affect significantly the obtained
mean kinetic energy.

\section{Conclusions}

In this paper we developed the exact formalism that describes the NCPs
obtained from DINS experiments in terms of an integral equation in the
variable $y$ that involves a kernel dependent on the experimental
setup parameters. For its calculation, a precise knowledge of the
detectors efficiency, the filter total cross section and the incident
neutron spectrum in a wide range of energies is required. As a result,
we obtained a Volterra equation of the first kind that allows to
obtain the desired momentum distribution $J(y)$ by means of numerical
methods.

A significant part of this paper was devoted to the study of the
variable $y$, that in the exact formalism here presented has a
definition that differs significantly from that employed in the CA
framework. The physical significance of this variable is the
projection of the initial impulse of the target nuclei over the
direction of the momentum transfer vector $\bm {q}$. In DINS data
processing procedures the passage from $t$ to $y$ is customarily
calculated taking a fixed final energy, and thus it has a single
well-defined value for each time channel. In this paper it was
calculated exactly. The importance of knowing the exact mapping of the
variable $y$ in the $(E_0,E)$-plane can be referred to our recent
study on the neutron final energy distributions in DINS experiments
\cite{FinalE}. In that work we found that such distributions are far
more complex than considering a single final energy, and depends on
the time channel and the dynamics of the scattering species. We showed
the constant-$y$ curves on the $(E_0,E)$-plane for hydrogen and
deuterium at a particular scattering angle and ratio of flight paths,
which served to illustrate its behavior, and developed its general
expressions. A noticeable particularity of the $y$ variable in the
exact treatment that we present in this paper, is that it can have
local extrema as a function of $E_0$ for a given time of flight as
shown in Fig.  \ref{figye0}, which define straight lines in the
$(E_0,E)$-plane (LL).  This causes singularities in the Jacobian
$dE_0/dy_M$ which affect the kernel $f_M(y_M,t)$. For example in the
analyzed case of hydrogen, values of $y$ greater that that those
defined by the LL are not physically allowed.

In past publications \cite{Blos1,Blos2} we showed that CA can not be
deduced from the exact expression. In this paper we again showed that
the CA formalism is incompatible with the exact one, and that the
basis of the difference between both formalisms resides in the
integration kernels.  Thus, the resolution function must be replaced
by a kernel that depends on $t$.  In Fig.  \ref{figfyt} we show
explicitly the differences between the exact kernel and the commonly
accepted resolution function. Besides the clearly observed differences
in shape, it is notable the appearance of the above mentioned
singularities. In the particular case analyzed in this paper these
discrepancies are particularly noticeable at 74.91 $\mu$sec, where the
singularity occurs exactly at the value of $y$ where the maximum of
the resolution function is centered, {\it i.e. } at $y=\tilde y_M(t)$.

So far, we have focussed our discussion on the differences between the
kernel $f_M(y_M,t)$ and the resolution function, without making any
reference to the momentum distribution of the sample $J_M(y_M)$.  As a
first exam on the interrelation between $f_M(y_M,t)$ and $J_M(y_M)$ it is
helpful to check to what extent the singularity can affect significant
parts of the integrand (\ref{cdety2}). To this end we will compare the
ratio $E/E_0$ where the singularities appear (defined by the slope of
the LL) with the same ratio for the scattering on a sample defined by
an ideal gas at $T=$0K, {\it i.e.} \cite{Blos1}
\begin{equation}
  \label{fblos}
  \frac{E}{E_0}=\left(\frac{\cos\theta+\sqrt{\cos^2\theta+(A-1)(A+1)}}{A+1}\right)^2.
\end{equation}
Eq. (\ref{fblos}) defines also the most probable energy ratio in the
case of a sample at finite temperature. In Fig. \ref{figeth} we show
in open symbols the ratio $E/E_0$ of the LL for H, D and O, and in full
symbols that corresponding to Eq. (\ref{fblos}), as a function of the
scattering angle. We observe that for this particular value of $\ell$ the
singularities do not appear significantly near the main peak.  However,
due to the thermal motion (described by the shape of $J(y)$) the
singularities affect the description of the tails of NCPs.

The present formalism was formulated regardless of the function
$J(y)$. Although in this paper the case of an isotropic sample was
examined, all the considerations we developed are applicable also for
the case of anisotropic samples. It is also worth to notice that
although in most of the significant cases concerning the study of
Condensed Matter the function $J(y)$ is closely represented by a
Gaussian shape \cite{Mayers90}, this function could represent an
anisotropic momentum distribution, not centered in $y=0$ like in the
case of particles flowing in a preferential direction. With regard to
the analysis of experiments where the sample can be represented by the
usual Gaussian distributions, it must be emphasized that the effects
of the singularities will normally be smoothed out in the NCP, {\it
  i.e.} the experimentally accessible magnitude. However, the
differences between the exact formalism and the CA in the description
of the NCP are clearly manifested, as accounted in detail in Ref.
\onlinecite{Blos1}, so we will briefly summarize them here. A
comparison of the results produced by Eq. (\ref{convo}) in the
calculation of the NCPs, with the exact formalism represented either
by Eqs. (\ref{ct}) or (\ref{cdety1}) (both formulations are
equivalent, since they differ only in the integration variable),
reveal that the CA is defective in the description of
\begin{itemize}
\item the peak areas,
\item the peak centroids,
\item the peak widths,
\end{itemize}
being the case of light scatterers at low temperatures the most
sensitive ones to these imperfections. In this case the widths of the
momentum distribution function $J(y)$, and the resolution function
$R(y)$ becomes comparable, and thus the consequences of the
differences between $R(y)$ and the exact kernel is more clearly
manifested on the NCPs. Thus, the obtainment of peak areas and
effective temperatures is significantly affected because of the use of
the CA. The analysis shown in Sect. \ref{meankin}, clearly reflects
this assertion, and the inadequacy to employ resolution functions
(either theoretically or experimentally defined) that due to the
approximate nature of the CA has an unavoidable degree of ambiguity in
its definition.

We wish to conclude this paper giving a schematic outline of the
procedure we recommend to use to the experimentalists in the DINS
field. For the measurement of an impulse distribution function it is
necessary to characterize different parameters of the experimental
configuration.
\begin{enumerate}
\item The incident neutron energy spectrum can be determined by a
  measurement with a detector of well-characterized efficiency placed
  on the direct beam. 
\item The efficiency of the detectors bank can be characterized using
  a heavy target (like Pb or Bi) at the sample position (with the
  filter out of the scattered neutron path).
\item The effective filter thickness can be measured with a lead
  target, employing the method `filter in - filter out', and fitting
  it as a parameter in Eq. (\ref{ct}). Alternatively, a transmission
  experiment can be performed on the direct beam. In the process, the
  full total cross section of the filter is needed. A wide variety of
  cross sections can be found {\it v. gr.} in Ref. \onlinecite{ENDF}.
\item Experimental data must be corrected by multiple scattering and
  attenuation effects, using Monte Carlo procedures like those
  described in Refs. \onlinecite{MonteCarloEllos} and
  \onlinecite{MonteCarloNuestro}.
\item From the steps 1 to 3, which aim to know the kernel $f(y,t)$,
  and step 4 which corrects for finite sample effects, we obtain the
  necessary elements to state Eq. (\ref{cdety2}), that can be solved
  either by numerical means or by fitting parameters if we have a
  previous knowledge of the function $J(y)$. In this last instance the
  fitting process can be made also directly on the $t$ scale employing
  Eq. (\ref{ct}) without changing the integration variable from $E_0$
  to $y$.

\end{enumerate}

Finally, let us mention that the present formalism is also applicable to
the method of double differences recently presented \cite{dd}, which
consists in employing two filters of different thicknesses.  The use
of the formalism presented in our paper will be most beneficial for
the users community of these DINS techniques.

\section*{ACKNOWLEDGEMENTS}

This work was supported by ANPCyT (Argentina) under Project PICT No.
03-4122. We also thank Fundaci{\'o}n Antorchas (Argentina), and CLAF
(Centro La\-ti\-no\-a\-me\-ri\-ca\-no de F{\'\i}sica) for financial support.

\appendix
\section{Derivation of $dy/dE_0$}
\label{App}

From the definition of the $y$ of Eq. (\ref{y}) we have
\begin{eqnarray}
\label{dyde02}
\frac{dy}{dE_0}\Big|_{t}&=&\frac{\partial y}{\partial q}\frac{dq}{dE_0}\Big|_{t}+
\frac{\partial y}{\partial \omega}\frac{d\omega}{dE_0}\Big|_{t} \nonumber \\
&=&\left(\frac{-M\omega}{q^2}-\frac{\hbar}{2}\right)\frac{dq}{dE_0}\Big|_{t}+\frac{M}{q}\frac{d\omega}{dE_0}\Big|_{t}
\end{eqnarray}
So, we need to calculate the derivatives of $q$ and $\omega$ with respect
to $E_0$ at constant $t$.  Form Eq. (\ref{hw}) we have
\begin{equation}
\label{dwde0}
\frac{d \omega}{dE_0}\Big|_{t}=\frac{1}{\hbar}\left(1-\frac{dE}{dE_0}\Big|_{t}\right)
\end{equation}
From the kinematic condition (\ref{tof}) we have
\begin{equation}
\label{dEdE0}
\frac{dE}{dE_0}\Big|_{t}=-\frac{L_0}{L_1}{\left(\frac{E}{E_0}\right)}^{3/2}
\end{equation}
Replacing Eq.(\ref{dEdE0}) in Eq.(\ref{dwde0}) we have
\begin{equation}
\label{dwdE02}
\frac{d \omega}{dE_0}\Big|_{t}=\frac{1}{\hbar}\left(1+\frac{L_0}{L_1}\left(\frac{E}{E_0}\right)^{3/2}\right)
\end{equation}
On the other hand, deriving the definition of $q$ presented in Eq. (\ref{hq}) we have
\begin{equation}
\label{dqdE0}
\frac{dq}{dE_0}\Big|_{t}=\frac{\partial q}{\partial E_0}\Big|_{t}+\frac{\partial q}{\partial
  E}\frac{\partial E}{\partial E_0}\Big|_{t}.
\end{equation}
Then
\begin{equation}
\label{dqdE02}
\frac{dq}{dE_0}\Big|_{t}\!=\! 
\frac{m}{\hbar^{2}q}
\left[1-\frac{L_0}{L_1}{\left(\frac{E}{E_0}\right)}^{3/2} \!\!+ \!
\left(\frac{L_0}{L_1}\frac{E}{E_0}\!-\!\sqrt{\frac{E}{E_0}}\right)\cos\theta\right]
\end{equation}
where we have employed Eq.(\ref{dEdE0}). Finally, replacing
Eq. (\ref{dwdE02}) and Eq. (\ref{dqdE02}) in Eq. (\ref{dyde02}) we
obtain Eq. (\ref{dyde01}).

\section{Derivation of $E(E_0,y)$}
\label{App2}

From the Eq. (\ref{ydeE0}) for $y$ (which is in terms of $E$ and $E_0$), we want to solve it 
for $E$, considering $y$ and $E_0$ as fixed input values. Defining the variables

\begin{equation}
\label{adimen2}
\left\{
\begin{array}{lll} 
x&=&\sqrt{E} \nonumber \\
\alpha&=&\frac{\displaystyle M^2}{\displaystyle 2my^2} \nonumber \\
\mu&=&\frac{\displaystyle m}{\displaystyle M} \nonumber \\
\end{array}
\right.,
\end{equation}
after some algebra Eq. (\ref{ydeE0}) can be transformed to a
fourth-order polynomial in $x$, with real coefficients
\begin{equation}
\label{poli2}
\mathop{\sum}_{i=0}^{4}b_i{x^i}=0
\end{equation}
where
\begin{eqnarray}
\label{coef2}
b_0&=&\left[\alpha  \left(1-\mu \right)^2 E_0-1\right] E_0  \nonumber \\
b_1&=&2 \left[1+2\alpha \left(\mu-\mu^2\right) E_0\right] \sqrt{E_0} \cos\theta\nonumber \\
b_2&=&-1-2 \alpha \left[1-\mu^2-2 \mu^2 {\cos}^{2}\theta\right]E_{0}  \\
b_3&=&-4{\alpha} {\mu} \left(1+\mu\right)\sqrt{E_0}\cos\theta \nonumber  \\
b_4&=&\alpha\left(1+\mu\right)^2.  \nonumber
\end{eqnarray}
Therefore, the real positive roots of Eq. (\ref{poli2}) define the
values of $E(E_0,y)$, which define the $y$-constant curves represented
in Figs. \ref{figmap1} and \ref{figmap2}.

\end{document}